\documentclass[aps,prl,superscriptaddress,twocolumn,amsmath,amssymb,showpacs]{revtex4-1}

\usepackage[english]{babel}
\usepackage{times}
\usepackage{epsfig,graphicx}
\usepackage[tight]{units}

\newcommand{\comment}[1]{}
\usepackage{color}

\begin{document}

\title{Spatio-temporal dynamics of bumblebees foraging under predation risk}

\author{Friedrich Lenz}
\affiliation{School of Mathematical Sciences, Queen Mary University of London, Mile End Road, London E1 4NS, UK}

\author{Thomas C. Ings}
\affiliation{School of Biological and Chemical Sciences, Queen Mary University of London, Mile End Road, London E1 4NS, UK}

\author{Lars Chittka}
\affiliation{School of Biological and Chemical Sciences, Queen Mary University of London, Mile End Road, London E1 4NS, UK}

\author{Aleksei V. Chechkin}
\affiliation{Institute for Theoretical Physics, NSC KIPT, ul.\ Akademicheskaya 1,UA-61108 Kharkov, Ukraine}

\author{Rainer Klages}
\email{r.klages@qmul.ac.uk}
\affiliation{School of Mathematical Sciences, Queen Mary University of London, Mile End Road, London E1 4NS, UK}

\begin{abstract}
We analyze 3D flight paths of bumblebees
searching for nectar in a laboratory experiment with and without
predation risk from artificial spiders. For the flight velocities we
find mixed probability distributions reflecting the access to the
food sources while the threat posed by the spiders shows up only in
the velocity correlations. The bumblebees thus adjust their flight
patterns spatially to the environment and temporally to predation
risk. Key information on response to environmental changes is
contained in temporal correlation functions, as we explain by
a simple emergent model.
\end{abstract}



\maketitle

  Quantifying foraging behavior of organisms by statistical analysis has
  raised the question whether biologically relevant search strategies
  can be identified by mathematical modeling
  \cite{Pyke,Schick2008,Bartumeus2009,Luz2009,smouse,Benichou2011review,Viswanathan2011}.
  For sparsely, randomly distributed, replenishing food sources, the
  L\'{e}vy flight hypothesis predicts that a random search with jump
  lengths following a power law minimizes the search time
  \cite{LWandLFShlesingerKlafter,ViswanathanStanley1999,Viswanathan2011}.  Experimental
  evidence
  \cite{Viswanathan1996,ReynoldsDisplacedHoneybees,Sims2008,Sims2010}
  and further theoretical analyses
  \cite{Lomholt2008,Reynolds2009-Levy-predicted-to-be-emergent}
  supporting this hypothesis were challenged by refined statistical data
  analyses
  \cite{Edwards2007,Edwards2008,Dieterich2008,SimonaHapca01062009} and
  more detailed theoretical modeling
  \cite{SimsPitchfordMinimizingErrorsLevy,pitchford-efficient-or-inaccurate,Benichou2011review}.
  A crucial problem is how dispositions of a forager like memory
  \cite{Burns2005} or sensory perception \cite{SpaetheTautzChittka}, as
  well as properties of the environment
  \cite{Bartumeus2003,Morales2004,Bartumeus2005,Sims2008,Kai2009,Sims2010}, can be
  tested in a statistical foraging analysis
  \cite{Pyke,Schick2008,Bartumeus2009,smouse,Viswanathan2011}.
  Especially for data obtained from foraging experiments in the wild,
  it is typically not clear to which extent extracted search patterns
  are determined by forager dispositions, or reflect an adjustment of
  the dynamics of organisms to the distribution of food sources and the
  presence of predators
  \cite{Sims2008,Sims2010,smouse}. This problem can be addressed by
  statistically quantifying search behavior in laboratory experiments
  where foraging conditions are varied in a fully controlled manner
  \cite{Bartumeus2003,Sims2010}.  Such an experiment has been
  performed by Ings and Chittka
  \cite{Ings2008,Ings2009}, who studied the foraging behavior of bumblebees 
  with and without different types of artificial spiders mimicking
  predators.

  Here we ask the question whether changes of environmental
  conditions as performed in the experiment by Ings and Chittka lead to
  changes in the foraging process.
  We answer this question by a statistical analysis of the
  bumblebee flights recorded in this experiment on both spatial and
  temporal scales. For this purpose, we extract both flight velocity
  probability distributions and temporal velocity autocorrelation
  functions from the data. Surprisingly, we find that the crucial
  quantity to understand changes in the bumblebee dynamics under
  predation risk is not the velocity distribution
  but the velocity correlation function, which reveals
  non-trivial dynamics on different time scales. We reproduce
  these changes by a simple Langevin equation modeling a repulsive
  interaction between insect and predator.  In order to
  construct mathematical models reproducing the foraging of
  organisms that interact with the environment, our results suggest to
  shift the focus from scale-free approaches
  \cite{ViswanathanStanley1999,Viswanathan1996,Viswanathan2011} to
  the statistical quantification of spatio-temporal changes in the
  foraging dynamics.

 \begin{figure}
    \includegraphics[scale=0.65]{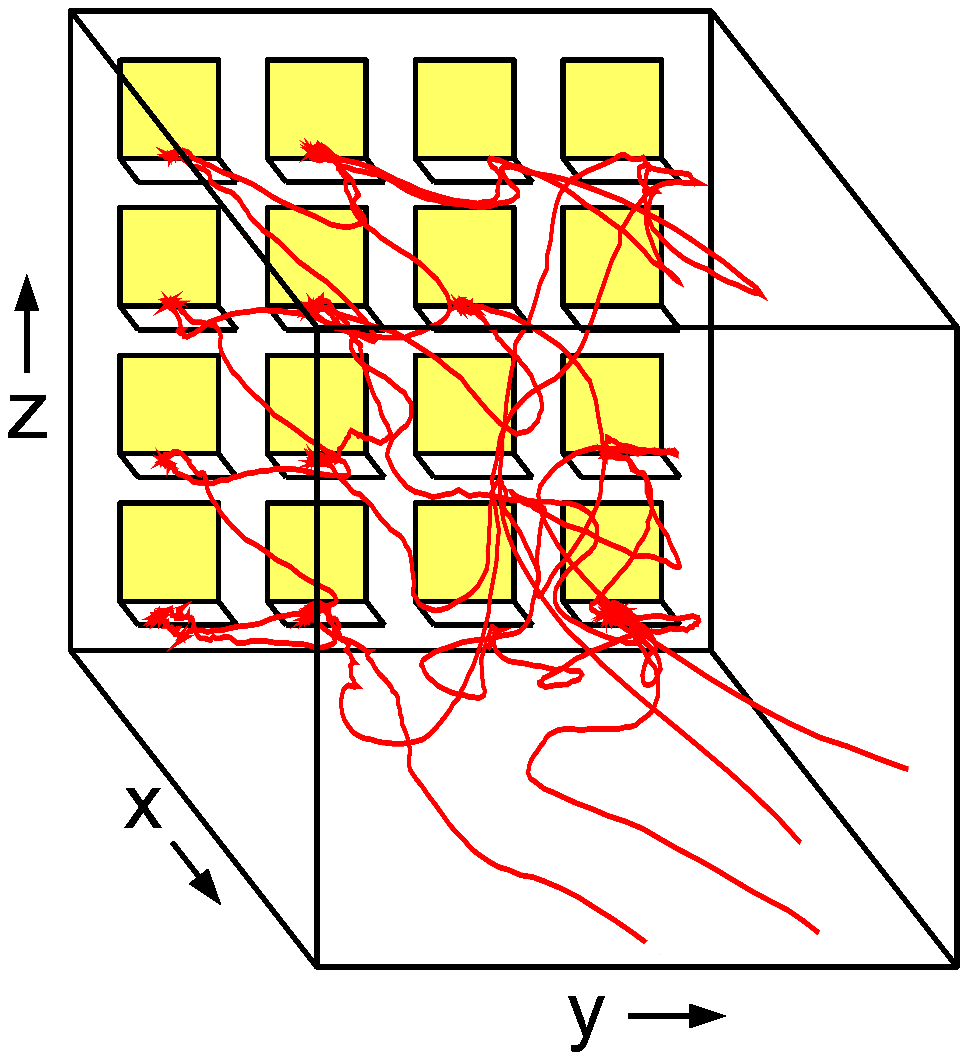}
    \caption{\label{fig:cage}
      Diagram of the foraging arena together with part of the flight trajectory of
      a single bumblebee.  
      The bumblebees forage on a grid of artificial flowers on one wall of the
      box. While being on the landing platforms, the bumblebees have access to
      food supply.  All flowers can be equipped with spider models and trapping
      mechanisms simulating predation attempts.
    }
  \end{figure}

  In the experiment \cite{Ings2008} bumblebees ({\it Bombus
  terrestris}) were flying in a cubic arena of $\approx \unit[75]{cm}$
  side length by foraging on a 4$\times$4 vertical grid of artificial
  yellow flowers on one wall. The 3D flight trajectories of 30
  bumblebees, tested sequentially and individually, were tracked by two
  high frame rate cameras ($\Delta t =
  \unit[0.02]{s}$).  On the landing platform of each flower, nectar was
  given to the bumblebees and replenished after consumption.  The
  short trajectory in Fig.~\ref{fig:cage}
  shows a typical flight path of a bumblebee foraging in the arena.
 To analyze differences in the foraging behavior of the bumblebees
  under threat of predation, artificial spiders were introduced.
  The experiment was staged into three phases: (1) spider-free foraging, (2)
  foraging under predation risk and (3) a memory test one day later.
  Before and directly after stage (2) the bumblebees were trained to forage in
  the presence of artificial spiders, which were randomly placed on 25\% of the
  flowers.
  A spider was emulated by a spider model on the flower and a trapping mechanism
  which held the bumblebee for two seconds to simulate a predation attempt.
  In (2) and (3) the spiders models were present but the traps were inactive in order
  to analyze the influence of previous experience with predation risk on the
  bumblebees' flight dynamics.
  To determine whether the detectability of the spiders is an important factor, half
  of the bumblebees were trained on easily visible (white) spider models and
  half of them on yellow models, which meant that spiders
  were camouflaged on the yellow
  flowers; see Ings and Chittka \cite{Ings2008} for further details of the
  experiment.

  \begin{figure}
    \includegraphics[scale=0.3,angle=-90]{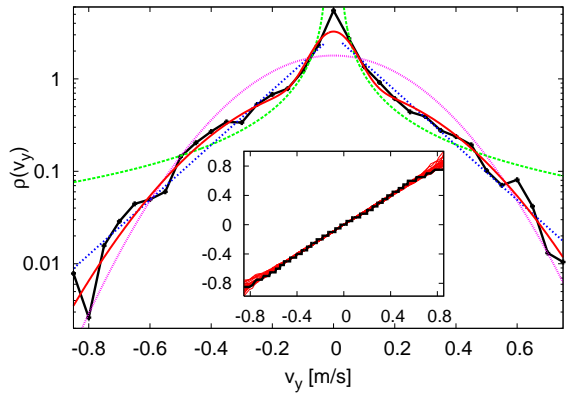}
    \caption{ \label{fig:qq}
      Estimated velocity distributions (main part) and Quantile-Quantile probability
			plot of a Gaussian mixture as the best fit (inset).
      Semi-logarithmic plot of the normalized histogram of velocities $v_y$
      parallel to the $y$-axis in Fig.~\ref{fig:cage} (black crosses) for a single
      bumblebee in the spider-free stage (1) together with a Gaussian mixture (red
      line), exponential (blue dotted), power law (green dashed), and Gaussian
      distribution (violet dotted), fitted via maximum likelihood
      estimation.\label{fig:histo} The inset shows quantiles of $v_y$ (in
      $\unitfrac{m}{s}$) of a single bumblebee against quantiles of an estimated
      mixture of two Gaussians. An ideal match would yield a straight line. The
      dashed red lines show 20 surrogate data sets of the same size.
    }
  \end{figure}

  Figure \ref{fig:histo} shows a typical normalized histogram of the horizontal
  velocities parallel to the flower wall (cf.\ y-direction in
  Fig.~\ref{fig:cage}) for a single bumblebee.
  The histograms are characterized by a peak at low velocities and vary in the
  different spatial directions due to asymmetries induced by physical and
  biological constraints as well as the spatial arrangement of the flowers.
  Direct fitting of distributions on the histogram and a visual comparison with
  some assumed distribution was shown to be unreliable \cite{Edwards2008}, as is
  illustrated by Fig.~\ref{fig:histo}: only the power law and the Gaussian
  distribution can be ruled out by visual inspection.
  However, the Gaussian mixture and an exponential function appear to be equally
  likely.
  Therefore we use the maximum likelihood method for a number of candidate
  distributions to obtain the optimal parameters for each candidate and then
  compare the different distribution types by their weights using the Akaike
  information criterion \cite{Edwards2007}.
  Our candidate distributions are:
  (a) Exponential: $\rho_\lambda(v) = c e^{-\lambda \left|v\right|}$,
  (b) Power law: $\rho_\mu(v) = c \left|v\right|^{-\mu}$,
  (c) Normal distribution with zero mean:
    $\rho_{\sigma}(v) = N_\sigma(v)$,
  (d) Mixture of two normal distributions:
    $\rho_{a,\sigma_1, \sigma_2}(v) = a N_{\sigma_1}(v)+(1-a)N_{\sigma_2}(v)$,
  where $N_{\sigma_i}(v)= \frac{1}{\sqrt{2\pi\sigma_i^2}}e^{-\frac{v^2}{2\sigma_i^2}}\,,\,i=1,2$,
  and $0\le a\le1$. Details of this analysis are described in the Supplemental Material~\cite{supplement}.

  For the data sets of all bumblebees and in all stages of the
  experiment the Akaike weights show that a mixture of two Gaussians
  is the preferred distribution of the tested candidates (see Table I 
  in the Supplemental Material \cite{supplement}).
  However, they do not inform us if the best of the candidates is actually a
  good model: if all of the candidates are far off the real distribution, the
  Akaike weights could highlight one of them as the best of the poor fits.
  As a supplementary qualitative test to which extent the estimated distribution
  with the largest Akaike weight deviates from the data over the whole range
  variables, we use Quantile-Quantile (Q-Q) probability plots.
  The inset of Fig.~\ref{fig:qq} shows the Q-Q plot of the mixture of two
  Gaussians against the experimental data of a single bumblebee and 20 surrogate
  data sets.
  Each of the surrogate data sets consists of independently identically
  distributed random numbers drawn from the estimated Gaussian mixture and has
  the same number of data points as the real data for comparison with
  statistical fluctuations.
  The Q-Q plot shows that the deviations of the experimental data from the
  mixture of two Gaussians is not larger than the expected deviations due to the
  finite quantity of data.

  The Gaussian mixture for the velocities is generated by different flight
  behavior near a flower versus in open space, which bears
  some
  resemblance to intermittent dynamics \cite{Morales2004,Benichou2011review,Viswanathan2011}.
  This has been verified by splitting the data into flights far from the flower
  wall vs.\ flights in the feeding zone.
  The latter was defined by a cube of side length \unit[9]{cm} around each
  flower in which the velocities are determined by approaching a flower and
  hovering behavior.
  This separation of different flight phases is thus adapted to accessing the
  food sources and explains the origin of Gaussian distributions with different
  variances in both spatial regions. 
  Because of the absence of a sparse distribution of food sources,
  there was no reason to expect L\'{e}vy-type probability
  distributions \cite{ViswanathanStanley1999}.
  Surprisingly, by comparing the best fits to these distributions for the
  different stages of the experiment, we could not detect any differences in the
  velocity distributions between the spider-free stage and the stages where
  artificial spider models were present, as is shown in Table II 
  of the Supplemental Material  \cite{supplement}.
  The parameters of the Gaussian mixture vary between individual bumblebees but
  there is no systematic change due to the presence of predators.

\begin{figure}
    \includegraphics[scale=0.45]{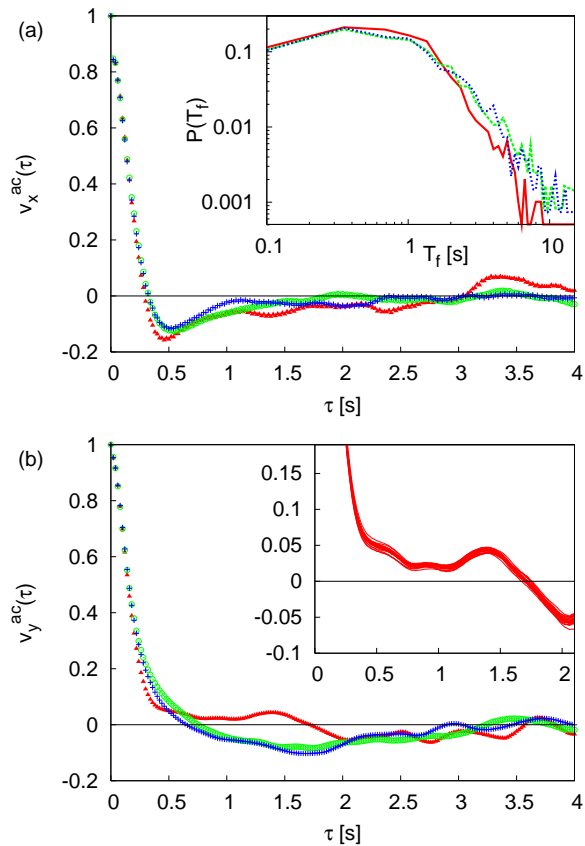}
    \caption{ \label{fig:corr}
      Autocorrelation of the velocities at different experimental stages:
      without spiders (red triangles), under threat of predation (green circles), and under threat a
      day after the last encounter with the spiders (blue crosses).
      (a)~In the x-direction perpendicular to the wall the velocities
      are anti-correlated for small times
      ($\approx \unit[0.5]{s}$) due to short flights from one flower to a nearby
      flower back at the flower wall. Inset: the distribution of flight times
      $T_f$ for each stage shows a corresponding maximum for these short jumps.
      Under threat of predation (dotted) long flights become more frequent.
      (b)~The correlation of $v_y$ parallel to the wall shows the
      effect of the presence of spiders on
      the flight behavior of the bumblebees.  The inset shows the resampled
      autocorrelation for the spider-free stage in the region where the
      correlation differs from the stages with spider models, which confirms that
      the positive autocorrelations are not a numerical artifact.
    } 
  \end{figure}

  Hence, we examined the velocity autocorrelations for complete flights
  from flower to flower. The autocorrelations have been computed by
  averaging over all bumblebees while weighting with the amount of
  data available for each time interval; see the Supplemental
  Material for details~\cite{supplement}.
  Figure \ref{fig:corr} shows the velocity autocorrelations in the x- and
  y-directions for different stages of the experiment.
  In the x-direction perpendicular to the wall the velocities are
  always anti-correlated for times around $\unit[0.5]{s}$
  (Fig.~\ref{fig:corr}(a)), which is due to the tendency of the
  bumblebees to quickly return to the flower wall.
  However, the flights with long durations between flower visits become more
  frequent for stages (2) and (3) where the bumblebees were exposed to
  predation risk compared with stage (1) (inset of
  Fig.~\ref{fig:corr}(a)).
  This is also reflected in a small shift of the global minimum
  in the correlations for stages (2) and (3) away from the origin.
  The $v_x$-autocorrelations thus display similar functional
  forms but with quantitative changes between the different
  stages. In contrast, for the $v_y$-autocorrelations the functional
  forms change profoundly: Parallel to the flower wall the velocities
  are anti-correlated in the presence of spiders for
  $\unit[0.7]{s}<\tau<\unit[2.8]{s}$, while for the spider-free stage
  the correlations remain positive up to $\unit[1.7]{s}$
  (Fig.~\ref{fig:corr}(b)).
  The vertical z-direction is similar to the y-direction with a weaker
  dependence on the presence of predators.
  As the limited amount of data causes variations in the
  autocorrelations between individual bumblebees, we resampled the
  result by leaving the data of each single bumblebee out (jackknifing).
  The resampling (inset of Fig.~\ref{fig:corr}(b)) confirms that
  the positive autocorrelations of $v_y$ are not a numerical artifact.

  Our statistical analysis of the experimental data has thus revealed
  that differences in the foraging behavior of bumblebees, triggered
  by predation risk, show up in changes of the velocity
  autocorrelation functions only, and not in modifications of the
  velocity probability distributions. 
  These changes are consistent with a more careful search:
  When no threat of predators is present, the bumblebees forage more
  systematically with more or less direct flights from flower to flower, arching
  away from the flower wall.
  Under threat the trajectories become longer and the bumblebees
  change their direction more often in their search for food sources,
  rejecting flowers with spiders, as is supported by Fig.~\ref{fig:corr}.
  Further analysis rules out that the main
  features of the correlation functions are induced by the geometry of
  the experiment: In Fig.~\ref{fig:corr}(a), all flight time
  distributions display maxima around $T_f\approx\unit[0.5]{s}$ suggesting
  that times
  below $\simeq\unit[2]{s}$ are primarily related to flights between
  flowers. Boundary effects are only evident for flight times
  that fall within the tail of the distributions. The
  anti-correlations in the $y$- and  $z$-directions
  parallel to the flower wall thus cannot be induced by the walls but
  are generated by a reversal of directions at flowers under predatory
  threat.  For the $x$-direction, the return to the flower wall is
  responsible for the anti-correlation at small delay times,
  not the opposite wall, which is too far away to have a significant
  effect. Another interesting aspect is
  that 
	Ings and Chittka \cite{Ings2008} report that bumblebees increase the time they spend
	inspecting flowers bearing camouflaged spiders compared to conspicuous ones.
	\comment{Ings and Chittka \cite{Ings2008} report an extension of the time span
  bumblebees spent in the feeding zone when exposed to camouflaged spiders
  compared to conspicuous ones.}
	However, we did not detect any change in velocity distributions
	or autocorrelations in this case, which suggests that the bumblebees
	perform longer localized inspection flights without changing their velocities.

	In order to understand the changes shown in
	Fig.~\ref{fig:corr}, we model the dynamics of $v_y$ by the
	Langevin equation
	$$ \frac{dv_y}{dt}(t) = - \eta v_y(t) - \frac{\partial U}{\partial y}(y(t)) + \xi(t)\:, $$
	where $\eta$ is a friction coefficient and  $\xi$ Gaussian white noise.
	The potential $U$ mimics an interaction between
	bumblebee and spider. Specific data analysis shows that this
	force is repulsive and dominates any hovering behavior in the
	velocity correlation decay. Computer simulations of the above
	equation reproduce a change from positive to
	anti-correlations by increasing the repulsive force.
	Details are discussed in the Supplemental
	Material~\cite{supplement}.  Note that the correlation decay
	displayed in Fig.~\ref{fig:corr} rules out a mathematical
	modeling in terms of ordinary correlated random walks or
	L\'evy walks, which predict velocity correlations to decay
	strictly exponentially \cite{Bartumeus2005,Viswanathan2011}
	or algebraically \cite{GNZ85,ZuKl93b}, respectively.
	
  We emphasize that the experiment analyzed in this
  Letter does not match the conditions of the L\'evy flight
  hypothesis \cite{ViswanathanStanley1999}.
  L\'evy flights and L\'evy walks predict scale-free
  probability distributions \cite{Viswanathan2011} and generate
  trivial functional forms for the velocity correlations
  \cite{GNZ85,ZuKl93b}.
  Accordingly, experiments testing this hypothesis have
  focused on probability distributions, not on correlation decay
  \cite{Viswanathan1996,ReynoldsDisplacedHoneybees,Sims2008,Sims2010}.
  However, our results demonstrate that velocity
  autocorrelations can contain crucial information for understanding
  foraging dynamics, here in the form of a highly non-trivial
  correlation decay emerging from an interaction between
  forager and predator. Identifying such an emergent property
  in contrast to adaptive behavior, as we do with our simple model,
  has been highlighted as a crucial problem in
  foraging dynamics \cite{Sims2010}.  In addition, we observe a
  spatial variation of the velocity distributions. These findings
  illustrate the presence of different flight modes
  governing the foraging dynamics on different scales of time and
  space. Our results thus indicate
  that taking scale-free distributions as a paradigm beyond the
  conditions of validity of the L\'evy flight hypothesis might be 
  too restrictive an approach in
  order to capture complex foraging dynamics. A variety of
  mechanisms may naturally lead to different foraging dynamics on
  different length and time scales, e.g.,
  individuality of animals \cite{SimonaHapca01062009,Petrovskii,Hawkes}, an
  intermittent switching between quasi-ballistic persistent dynamics and
  localized search modes \cite{Dieterich2008,Benichou2011review}, or quantities
  over which one has averaged like time of day \cite{Sims2010}.
  As ignoring these mechanisms can lead to spurious power laws
  \cite{Edwards2007,Edwards2008}, it is important
  to look for the reasons of the occurrence of non-trivial distributions like
  mixtures, e.g., animals switching between different search modes.
  These mixtures may not always be optimal distributions for a particular search
  problem, but they are easy to produce, composable and flexible enough such
  that differences to some optimal distribution might not be large enough to
  give rise to evolutionary pressure \cite{Luz2009}.

  In summary, the fundamental question `What is the mathematically most
  efficient search strategy of foraging organisms?'  has, under specific
  conditions \cite{pitchford-efficient-or-inaccurate}, been answered by the
  L\'{e}vy flight hypothesis \cite{LWandLFShlesingerKlafter,ViswanathanStanley1999}.
  This question is well-posed under precise foraging conditions and has the big
  advantage that it is amenable to mathematical analysis.
  However, it does not capture the full complexity of a biological foraging
  problem \cite{Viswanathan2011}, which incorporates both the dependence of foraging on `internal'
  conditions of a forager (sensory perception \cite{SpaetheTautzChittka}, memory
  \cite{Burns2005}, individuality \cite{SimonaHapca01062009,Petrovskii,Hawkes})
  as well as `external' environmental constraints (distribution of food sources
  \cite{Sims2008,Sims2010,Kai2009}, day-night cycle \cite{Sims2010}, predators
  \cite{Ings2008,Ings2009}).
  Asking about the range of applicability of the L\'{e}vy flight hypothesis
  leads to the over-arching question `How can we statistically quantify changes
  in foraging dynamics due to interactions with the environment?', which
  requires to identify suitable measurable quantities characterizing such
  changes.
  This question highlights the need to better understand, and more carefully
  analyze, the interplay between forager and environment, which will yield
  crucial information for constructing more general mathematical foraging
  models.

  We thank Holger Kantz for valuable support and Nicholas W.\ Watkins as well as
  Michael F.\ Shlesinger for helpful comments.
  Financial support by the EPSRC for a small grant within the framework of the
  QMUL {\em Bridging The Gap} initiative is gratefully acknowledged.  

\comment{
  \section*{Author Contributions}
    L.C.\ and T.C.I.\ designed the experiment, T.C.I.\ performed the experiment,
    R.K.\ initiated the project by assembling the team, all authors designed the
    data analysis, F.L.\ performed the data analysis, F.L.\ and R.K.\ wrote the
    paper and all authors contributed to subsequent drafts.
}



%

\end{document}


\section{Supplemental Material}
  
\vspace*{-0.2cm}
\noindent {\bf 1. Statistical data analysis}\\[-2ex]

  To capture bumblebee flights only, we exclude any crawling
  behavior on the landing platforms by also removing all data
  within a \unit[1]{cm} boundary region of each platform.
  The size of this boundary is based on the size of the bumblebees,
  which have a height of approximately \unit[1]{cm}. While smaller
  cutoffs would not exclude all crawling behavior, the cutoff can be
  increased robustly within reasonable bounds. We have checked that,
  e.g.\ a \unit[2]{cm} cutoff does not have any influence on any of the
  analyzed quantities, as the amount of the data which would be
  excluded in addition is very small. This leaves
	from 2000 to 15000 data points (average: 6000) per bumblebee for each stage.
  We select the best model for the velocity distributions by maximum likelihood
  estimation and Akaike and Bayesian weights for our candidate distributions [16] 
  for $|v| \geq \unitfrac[2.5]{cm}{s}$.
  Given a set of measured velocities $D=\{v_1,v_2,...,v_n\}$ and a probability
  density function $\rho_{\lambda}(v)$, where $\lambda$ is a vector of k
  parameters, the {\it log-likelihood} of the probability density function for a
  finite resolution of the data ($\Delta v=\unitfrac[5]{cm}{s}$) simplifies to
  %
  $$\ln L(\lambda|D)=\!\sum_{v_j \in D}\ln P_{\lambda}(v_j)=\!\sum_{b \in \mathit{bins}}h[b]\ln\!\int_{min(b)}^{max(b)}\!\rho_{\lambda}(v)dv$$
  where $h(b)$ is the observed frequency in bin $b$.

  For each candidate distribution $\rho^i_{\lambda_i}$, $i\in\{1,2,3\}$, we
  maximize the log-likelihood $\ln L_i$ w.r.t. $\lambda_i$ locally with a
  Nelder-Mead algorithm by using a Monte Carlo method to find the global
  maximum.
  To find the preference between the different model distributions whose
  likelihoods $L_i$ are maximized at $\lambda_i^{max}$ the information criteria
  are $$IC_i= -2 \ln(L_i(\lambda_i^{max}|D))+s(n) k_i$$ with $s(n)=2$ for the
  Akaike information criterion and $s(n)=\ln(n)$ for the Bayesian information
  criterion as a penalty on the number of parameters $k_i$.
  The best model, denoted by $*$, is the one which minimizes the information
  criterion $IC_{*} = \min\limits_i(IC_i)$. The Akaike/Bayesian weights then
  give the preference of each model over the others as a probability
  $$w_i=\alpha \mathrm e^{-(IC_i-IC_{*})/2} \;,$$
  where $\alpha$ normalizes the weights to $\sum_i w_i=1$.

  The choice of the information criterion makes no strong difference for the
  model selection in this experiment.
  With the Akaike information criterion the Gaussian mixture is chosen with a
  weight of over 95\% for all bumblebees and all experimental stages.
  The Bayesian information criterion agrees with the Akaike information
  criterion on 90\% of all data sets. For the other 10\% it prefers a single
  Gaussian or an exponential distribution - these data sets turned out to be
  those with the least amount of data available.

  To compute the autocorrelation function $v^{ac}(\tau)$ of the flight velocities
  $$ v^{ac}(\tau)=\frac{\left< (v(t)-\mu)(v(t+\tau)-\mu) \right>}{\sigma^2} $$ 
  we average over all bumblebees and over time in all flights that are
  complete from starting on one flower to landing on the next.
  We exclude flights containing gaps and correlation terms, where in-between
  time $t$ and $t+\tau$ a flower was visited.

  \onecolumngrid

  \begin{table}[h]
    \caption{\label{table:params}Model weights and estimated parameters.
      Akaike and Bayesian weights both give preference to the mixture of two
      Gaussians for $v_y$ for most of the bumblebees.
      The weights are estimated individually and their mean and standard deviation
      (in brackets) are shown. The distribution parameters are also estimated
      individually for each bumblebee in each stage.
    }
    \begin{ruledtabular}
      \begin{tabular*}{\hsize}{@{\extracolsep{\fill}}lcccccc}
        Model: & (a) Exponential & (b) Power law & (c) Gaussian & \multicolumn{3}{c}{(d) Gaussian Mixture} \\
        \hline
        Akaike weight & 0.00 (0.00) & 0.00 (0.00) &
             0.04 (0.19) & \multicolumn{3}{c}{ 0.96 (0.19) } \\
        Bayesian weight & 0.04 (0.18) & 0.00 (0.00) &
             0.08 (0.26) & \multicolumn{3}{c}{ 0.88 (0.30) } \\
        Parameters & $\lambda$ & $\mu$ & $\sigma$ & $a$ & $\sigma_1$ & $\sigma_2 $ \\
        average (bumblebees) & 5.61 & 1.11 & 0.25 & 0.67 & 0.06 & 0.29 \\
        stddev (bumblebees) & 1.07 & 0.16 & 0.03 & 0.13 & 0.04 & 0.03 \\
      \end{tabular*}
    \end{ruledtabular}
  \end{table}
  
  \begin{table}[h]
    \caption{\label{table:comparestages} Weights and estimated parameters of the Gaussian mixture
      for the different experimental stages.
      Weights and parameters are estimated for each bumblebee.
      Shown are the mean over all individuals and the standard deviation (in
      brackets). The mixture of two Gaussians is the best fit in all stages. In the
      parameters of the distribution we observe no significant effect of the threat
      of predators on the bumblebees.
    }
    \begin{ruledtabular}
      \begin{tabular*}{\hsize}{@{\extracolsep{\fill}}lccccc}
         Stages & Akaike weight & Bayesian weight & $a$ & $\sigma_1$ & $\sigma_2$ \\
        \hline
        (1) Without spiders & 0.97 (0.15) & 0.93 (0.23) & 0.64 (0.11) & 0.06 (0.02) & 0.29 (0.03) \\
        (2) Under predation risk & 0.99 (0.04) & 0.90 (0.27) & 0.68 (0.13) & 0.06 (0.02) & 0.29 (0.02) \\
        (3) With risk, $1$ day later & 0.89 (0.29) & 0.80 (0.38) & 0.72 (0.16) & 0.07 (0.07) & 0.30 (0.03) \\
      \end{tabular*}
    \end{ruledtabular}
  \end{table}

\pagebreak
\twocolumngrid

	\begin{figure}[h]
  	\includegraphics[scale=0.35,angle=-90]{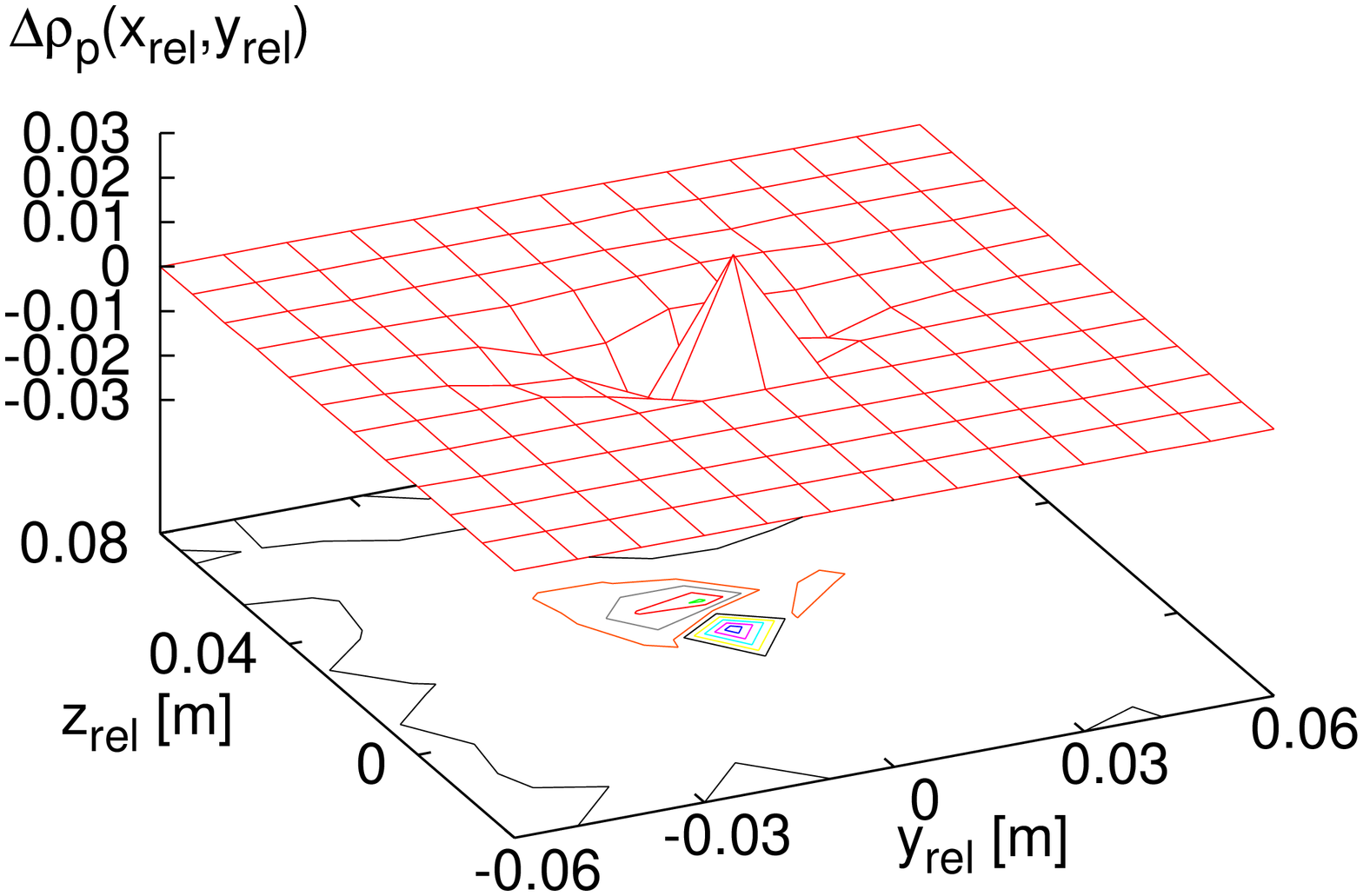}
  	\caption{\label{fig:floweravoidance}
	Predator avoidance of bumblebees at
	flowers, Eq.~(\ref{eq:pav}), extracted from the
	experimental data. Hovering
	behavior in front of a flower is represented by the positive
	spike directly at the flower center,
	while the negative region behind this spike reflects
	the avoidance in the flights towards a flower.
		}
	\end{figure}

\noindent {\bf 2. Mathematical modeling of bumblebee foraging}\\[-2ex]

	The effect of the presence of a spider on the probability of a
	bumblebee to fly in front of a flower can be measured by
	computing the difference between the position densities at
	stage (1) and (2) as a function of the positions
	parallel to and near $(x<$\unit[5]{cm}) the flower wall,
	\begin{equation}
	\Delta\rho_p(y_{rel},z_{rel})=\rho_p^{(2)}(y_{rel},z_{rel})-\rho_p^{(1)}(y_{rel},z_{rel})\:, \label{eq:pav}
	\end{equation} 
	where the positions $(y_{rel},z_{rel})$
  are relative to the nearest
	flower center. This predator avoidance extracted from
	the experimental data is shown in
	Fig.~\ref{fig:floweravoidance}.
	Two different types of behavior can be seen:
	First, there is a small increase in the amount of
	hovering, i.e.\ inspection flights near the flower platform
	when a spider model is present \cite{Ings2011,Fujisaki2008},
	which is consistent with Ref.~\cite{Ings2008}. However, more
	important is the local minimum representing the avoidance of
	flowers infected by spiders. This effect is strongest
	\unit[3]{cm} above the dangerous flowers, because the
	flowers are predominantly approached from above. The avoidance
	behavior affects not only flights
	near the flower wall but can still be detected further away from it.
	Comparing dangerous and safe flowers at stage (2) only
	confirms that avoidance is the dominant effect for
	search flights.
	
	The avoidance of spider-infected flowers together with the
	spatial switching of flight modes discussed in the main
	part of our Letter can be modeled by the Langevin Equation
	\begin{eqnarray} \label{eq:fullmodel}
	     \frac{d{\bf r}}{dt}(t) &=& {\bf v}(t)\nonumber\\
	   \frac{d{\bf v}}{dt}(t) &=& - \eta {\bf v}(t) - \nabla U({\bf r}(t)) + {\bf \xi}({\bf r},t)\:,
	\end{eqnarray}
	where $\eta$ is a friction coefficient and ${\bf \xi}$ white
	Gaussian noise with standard
	deviation depending on the flight mode as a function of the position,
	${\bf \xi}({\bf r},t)=\chi_{\mbox{fz}}({\bf
	r})\mathbf{\xi}_1(t)+(1-\chi_{\mbox{fz}}(\mathbf{r})){\bf
	\xi}_2(t)$.
	Here ${\bf r}=(x,y,z)^\top$ is the position of the bumblebee
	at time $t$, $\chi_{\mbox{fz}}({\bf r})$ is the indicator
	function of the feeding zone, which is equal to one whenever
	the bumblebee is in the cube around a flower as defined
	before, and $\mathbf{\xi}_i\,,\,i=1,2$ is Gaussian noise with
	two different variances. The potential $U$ models an
	interaction between bumblebee and spider in form of a
	repulsive force exerted by the spider onto the bumblebee,
  for which we assume that the potential maxima are located
  near infected flowers.

	When the mechanism generating the correlation functions shown
	in Fig.~3 is not the focus of the investigation, it suffices
	to consider a reduced version of Eqs.~(\ref{eq:fullmodel}) in
	form of the \emph{effective} Langevin equation
	\begin{equation}
	\frac{d{\bf r}}{dt} = \chi_{\mbox{fz}}({\bf r}){\bf\zeta}_1(t)
	+(1-\chi_{\mbox{fz}}({\bf r})){\bf \zeta}_2(t)\:.
	\end{equation}
	This equation describes the spatially varying hovering and
	search modes by using noise $\mathbf{\zeta}_i\,,\,i=1,2$.,
	which models the impact of the potential $U$ together with the
	noise ${\bf \xi}$.
	Further	data analysis shows that excluding hovering has no significant
	impact on the velocity autocorrelations, which are dominated
	by the search flights.  This is in full agreement with Fig.~3,
	where the time scale for the predator-induced anti-correlation
	(Fig.~3(b)) is larger than the time scale for flights between
	neighbouring flowers (Fig.~3(a)). Hence, we model
	${\bf \zeta_1}(t)$ as a vector of Gaussian white noise with the
	smaller variance $\sigma_1^2$ given in
	Table~\ref{table:params} which describes the hovering. The
	search flights from flower to flower are reproduced by the
	correlated Gaussian noise vector $\zeta_2(t)$ with variance
	$\sigma_2^2$ and the autocorrelations
	$v^{ac}_i(\tau)\:,\:i=x,y$ shown in Fig.~3. The advantage of
	this model is that it is directly based on our data analysis.

	We now focus on the different aspect of understanding the
	biophysical mechanism that
	generates the anti-correlations of the velocities parallel to
	$y$ shown in Fig.~3(b). Starting from the full model
	Eqs.~(\ref{eq:fullmodel}), it suffices to select
	the search mode only  by setting ${\bf \xi}({\bf r},t)={\bf
	\xi}_2(t)$ thus neglecting any spatial
	variations of the noise.  This yields the Langevin equation
	\begin{equation} 
	  \frac{dv_y}{dt}(t) = - \eta v_y(t) - \frac{\partial
	  U}{\partial y}(y(t)) + \xi(t)\label{eq:ule}\:,
	\end{equation}
	which was already stated in the main part as the main
	equation. A rough approximation for the repulsive force is
	provided by a periodic potential with maxima at dangerous
	flowers,
	\begin{equation}
	  U({\bf r}) = u \cos\left(2\pi \frac{y}{y_0}\right) \:,\label{eq:rep}
	\end{equation}
	where $y_0$ is the mean distance between spiders and $u$
	the strength of the repulsion.

\begin{figure}[t!]
	\includegraphics[scale=0.35,angle=-90]{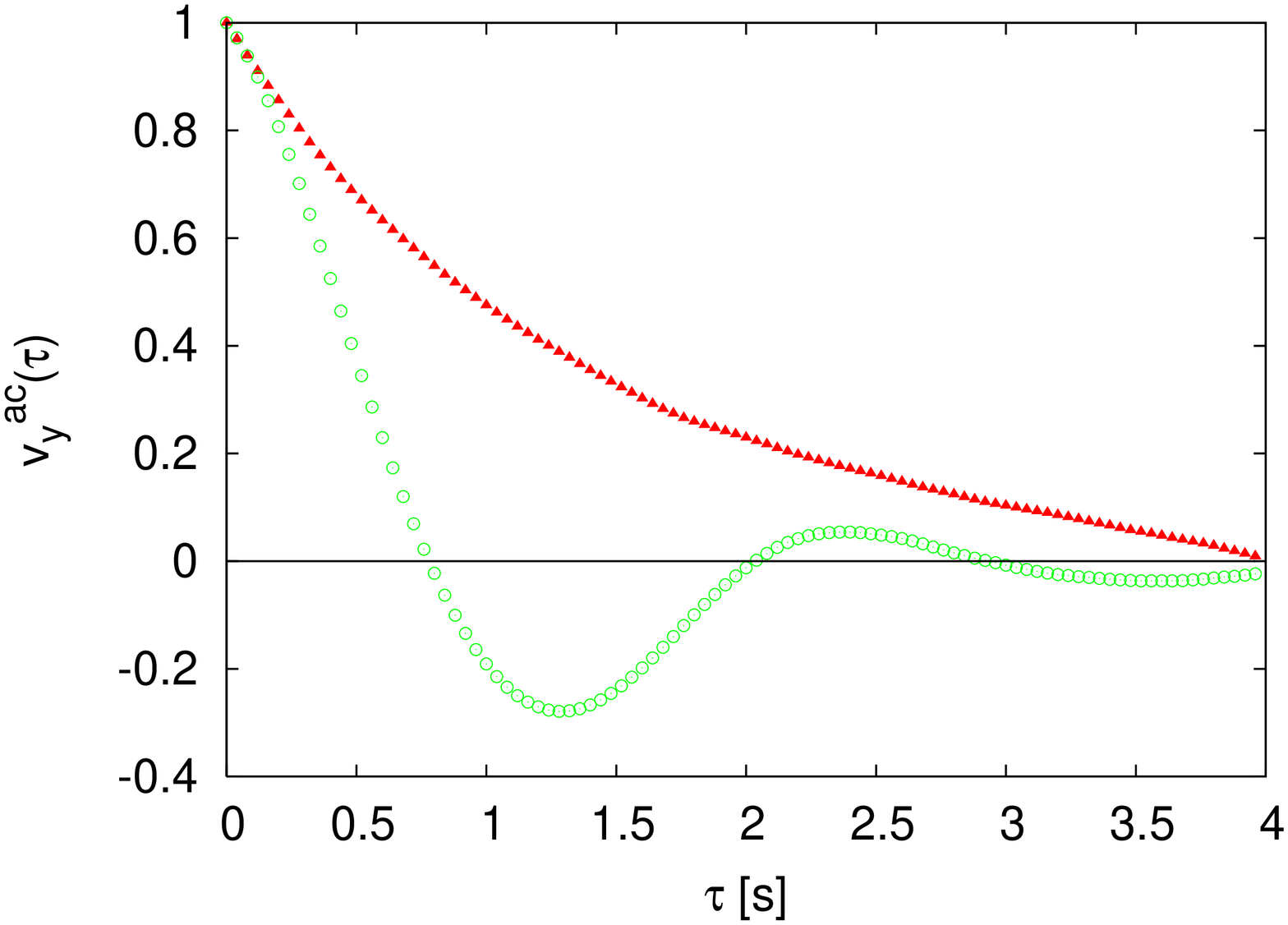}
	\caption{\label{fig:simulatedcorrelation} Autocorrelation
	function of the velocities $v_y$ for the Langevin model
	Eqs.~(\ref{eq:ule}),(\ref{eq:rep}) modeling predation threat
	by different strengths of a repulsive potential. Shown are
	results from computer simulations without ($u=0$; red
	triangles, upper line) and with predation threat ($u=\unit[0.5]{m^2/s^2}$; green
	circles, lower line). These results should be qualitatively
	compared with the experimental findings Fig.~3(b).}
	\end{figure}

	We integrated this Langevin equation via an Euler-Maruyama
	method under variation of $u$ by computing the autocorrelation
	function $v_y^{ac}$ of the generated
	data. Figure~\ref{fig:simulatedcorrelation} shows $v_y^{ac}$
	by increasing the repulsion strength $u$.
	The correlation function changes from positive correlations to
	anti-correlations in a range of delay times $\tau$ comparable
	to the changes in the correlation function of the experimental
	data of Fig.~3(b). This qualitatively reproduces our
	experimental findings from first principles. Note that the
	oscillations for higher $\tau$ in
	Fig.~\ref{fig:simulatedcorrelation} would
	be suppressed in a higher-dimensional model. The other
	directions can be treated analogously, e.g., by including
	an $x$-dependent term in the potential for the attraction of
	the bumblebees to the flower wall. A stochastic analysis of
	Langevin equations with periodic potentials can
	be found, e.g., in Ref.~\cite{Risken}.
	The effect of the harmonic potential on the creation of negative
  velocity correlations can also be calculated analytically \cite{ourlongpaper}.
	
	We emphasize that our model Eqs.~(\ref{eq:ule}),(\ref{eq:rep})
	provides only a qualitative description
	of the biophysical mechanism generating the change in the
	correlations of the bumblebee velocities under predation
	threat. For a quantitative comparison to the experimental data
	a much more detailed model would be necessary, which needs
	to include the random positioning of the spiders
	and the general attractive force exerted by the flowers onto
	the bumblebees. Modeling the three-dimensional nature of the
	potential would also be important: Notice, e.g., the local
	maximum of $v_y^{ac}$ around $\tau\simeq2.5$ which is an
	artifact of the one-dimensional modeling of spider avoidance.
	However, as it is difficult to reliably estimate the
	parameters of the potential, such a quantitative comparison is
	beyond the scope of our Letter.
	
\bibliographystyle{apsrev4-1}
\bibliography{all-copy.bib}
